# Reply to Comment on "Failure of the work-Hamiltonian connection for free-energy calculations" by Luca Peliti

Peliti's Comment [arXiv:0808.2855] improperly claims that our results in Ref. [1] contradict the well-established statistical mechanics expression of the free energy in terms of the partition function.

Our results never call into question that free energies are connected to the partition function. What they show is that parametric changes of the Hamiltonian cannot be connected in general to physical free energy changes. The distinction between energies and their changes is important. Energies, in general, are defined up to an arbitrary factor, which does not depend on the coordinates that describe the state of the system [2, p. 171]. The free energy difference of two states, in contrast, is a physical measurable quantity because the arbitrary factor is removed upon subtraction.

The problem with subtracting two free energies computed from Hamiltonians with different values of the parameters, as Peliti does, is that the arbitrary factor typically depends on the parameters [2, p. 171] and consequently the result obtained remains arbitrary. As a straightforward example, the Hamiltonian $H(x;g(t)) = H_0(x) + g(t)$, where $g(t)$ is a time-dependent parameter that does not affect the system, would lead to arbitrary free energy changes, $\Delta G_Z = g(t) - g(0)$, that do not result from any physical process but from the mathematical description [1].

Peliti's *im*proper *ad hoc* redefinition of work as $W_{IP} = \int dt \frac{\partial \lambda}{\partial t} \frac{\partial}{\partial \lambda} H(x(t); \lambda(t))$ has several unusual properties that guarantee the mathematical validity of Jarzynski equation [3], but it does not ensure the thermodynamic validity of the resulting free energy changes [1, 3, 4]. (Note that Peliti uses the notation $W$ for $W_{IP}$ and $W_0$ for $W = Force \times Displacement$.)

Peliti incorrectly states that $W_{IP}$ is a time-honored definition of work already used by Gibbs [5] and Tolman [6]. This misconception is mainly based on the confusion of external parameter with external coordinate. Gibbs and Tolman never used arbitrary parametric changes. When external coordinates are used, as Gibbs and Tolman did, changes of the Hamiltonian are the result of a force times a displacement.

To illustrate the extent of this confusion, let us consider the prototypical stretching of an RNA molecule with optical tweezers in the linear regime. The RNA molecule (with elastic constant $k$) is the system, the bead attached to the RNA molecule is the external body (described by the coordinate $x$), and the center of the optical trap (with force constant $K$) is the external time-dependent parameter (denoted by $X_t$). The work done by the system on the external body is $dW_{RNA} = -kxdx$ and the parametric change of the Hamiltonian is $dW_{IP} = -K(x - X_t)dX_t$ [1]. This example explicitly shows that $W_{IP}$, in contrast to what Peliti incorrectly assumes, is not the work done by the system on the external body, neither in magnitude nor in sign.

Thermodynamically, free energy changes are defined as the work $W$ performed over a reversible trajectory [5, p. 165; 6, p. 527]. For small systems, the work is averaged because it fluctuates [7, p. 1-26]. Indeed, the quantity that has been used experimentally to compute free energy changes of RNA molecules is the average of $W$ over reversible trajectories [8], not $W_{IP}$.

In summary, Peliti's Comment fails to appreciate basic physical principles and consequently misrepresents both our work as well as the classical work of Gibbs and Tolman.

J. M. G. Vilar[1] and J. M. Rubi[2]

[1]Computational Biology Program, Memorial Sloan-Kettering Cancer Center, New York, New York 10065, USA

[2]Departament de Fisica Fonamental, Universitat de Barcelona, Diagonal 647, 08028 Barcelona, Spain

**References**

[1] J. M. G. Vilar, and J. M. Rubi, Phys. Rev. Lett. **100**, 020601 (2008).
[2] D. Halliday, R. Resnick, and J. Walker, *Fundamentals of physics, 7th ed.* (Wiley, New York 2005).
[3] J. M. G. Vilar, and J. M. Rubi, cond-mat arXiv:0707.3802v1 (2007).
[4] B. Palmieri, and D. Ronis, Phys. Rev. E **75**, 011133 (2007).

[5] J. W. Gibbs, *Elementary principles in statistical mechanics* (Yale University Press, New Haven, 1902).
[6] R. C. Tolman, *The principles of statistical mechanics* (Oxford University Press, London, 1955).
[7] T. L. Hill, *Thermodynamics of small systems* (W.A. Benjamin, New York, 1963).
[8] J. Liphardt *et al.*, Science **292**, 733 (2001).